\documentclass[prl,showpacs,superscriptaddress,twocolumn]{revtex4}
\usepackage{graphics,dcolumn}
\graphicspath{{.}{./pictures/}}
\begin{document}
\preprint{nucl-ex/0102???}
\title{Neutral pion threshold production at
       {\boldmath $Q^2=0.05\,\mathrm{GeV^2/c^2}$}\\
       and Chiral Perturbation Theory}
\providecommand{\jogu}{Johannes Gutenberg-Universit\"{a}t Mainz, 
                       55099~Mainz, Germany}
\providecommand{\ikp}{\affiliation{Institut f\"{u}r Kernphysik, \jogu}}
\providecommand{\ipm}{\affiliation{Institut f\"{u}r Physik,      \jogu}}
\providecommand{\MIT}{\affiliation{Laboratory for Nuclear Science,  
                                   Massachusetts Institute of Technology, 
                                   Cambridge, MA~02139, U.S.A.}}
\providecommand{\ijs}{\affiliation{Jo\v{z}ef Stefan Institute, 
                                   SI-1001~Ljubljana, Slovenia}}
\providecommand{\tud}{\affiliation{Institut f\"{u}r Kernphysik, TU Darmstadt, 
                                   64289~Darmstadt, Germany}}
\providecommand{\ums}{\affiliation{Department of Physics, 
                                   University of Massachusetts,
                                   Amherst, MA~01003, U.S.A.}}
\author{H. Merkel}
\email{Merkel@kph.uni-mainz.de}
\homepage[\\URL: ]{http://wwwa1.kph.uni-mainz.de/}\ikp
\author{P.~Bartsch}\ikp
\author{D.~Baumann}\ikp
\author{J.~Bermuth}\ipm
\author{A.\,M.~Bernstein}\MIT
\author{K.~Bohinc}\ijs
\author{R.~B\"{o}hm}\ikp
\author{N.~Clawiter}\ikp
\author{S.~Derber}\ikp
\author{M.~Ding}\ikp
\author{M.\,O.~Distler}\ikp
\author{I.~Ewald}\ikp      
\author{J.\,M.~Friedrich}
\altaffiliation[Present address: ]{TU M\"unchen, Garching, Germany}\ikp
\author{J.~Friedrich}\ikp  
\author{P.~Jennewein}\ikp
\author{M.~Kahrau}\ikp 
\author{M.~Kohl}\tud
\author{K.\,W.~Krygier}\ikp 
\author{M.~Kuss}
\altaffiliation[Present address: ]{INFN Sezione di Pisa, Pisa, Italy}\tud
\author{A.~Liesenfeld}\ikp
\author{P.~Merle}\ikp      
\author{R.\,A.~Miskimen}\ums
\author{U.~M\"{u}ller}\ikp  
\author{R.~Neuhausen}\ikp  
\author{M.\,M.~Pavan}
\altaffiliation[Present address: ]{TRIUMF, Vancouver, B.C., Canada}\MIT
\author{Th.~Pospischil}\ikp
\author{M.~Potokar}\ijs    
\author{G.~Rosner}
\altaffiliation[Present address: ]{University of Glasgow, Glasgow, U.K.}\ikp
\author{H.~Schmieden}\ikp  
\author{M.~Seimetz}\ikp
\author{S.~\v{S}irca}
\altaffiliation[Present address: ]{M.I.T., Cambridge MA 02139, U.S.A.}\ijs
\author{A.~Wagner}\ikp
\author{Th.~Walcher}\ikp
\author{M.~Weis}\ikp
\collaboration{A1 Collaboration}\noaffiliation
\date{August 20, 2001}
\begin{abstract}
  New data are presented on the $p(e,e'p)\pi^0$ reaction at threshold at a
  four-momentum transfer of $Q^2=0.05\,\mathrm{GeV^2/c^2}$. The data were
  taken with the three-spectrometer setup of the A1 Collaboration at the Mainz
  Microtron MAMI. The complete center of mass solid angle was covered up to a
  center of mass energy of $4\,\mathrm{MeV}$ above threshold. Combined with
  measurements at three different values of the virtual photon polarization
  $\epsilon$, the structure functions $\sigma_T$, $\sigma_L$, $\sigma_{TT}$,
  and $\sigma_{TL}$ are determined. The results are compared with
  calculations in Heavy Baryon Chiral Perturbation Theory and with a
  phenomenological model. The measured cross section is significantly smaller
  than both predictions.
\end{abstract}
\maketitle
\paragraph*{Introduction.}

Threshold electromagnetic pion production is a fundamental process since the
pion is a ``Goldstone Boson'', reflecting the spontaneously broken chiral
symmetry of QCD\cite{wei79}. In the chiral limit, where the quark and pion
masses vanish, the $s$ wave production amplitudes of neutral pions
vanish. However, the 
explicit chiral symmetry breaking due to the small but finite quark mass
($m_{u}\approx 5\,\mathrm{MeV/c^{2}}$, $m_{d}\approx 9\,\mathrm{MeV/c^{2}}$)
and finite pion mass render these amplitudes finite.

Calculations of these observables are performed by an effective field theory
called Chiral Perturbation Theory (ChPT)\cite{wei79}, which is generally in
good agreement with experiment\cite{book}. The systematic application of
ChPT to reactions involving heavy baryons by Ref.~\cite{bernard} (Heavy Baryon
Chiral Perturbation Theory, HBChPT) has been generally successful in
describing $\pi-N$ scattering and electromagnetic pion production from the
nucleon.

In recent years there has been a considerable experimental effort to test this
theoretical approach\cite{book}. Of specific interest to this work, photo
production experiments were performed at Mainz\cite{Beck,Fuchs,Schmidt} and
at SAL\cite{Bergstrom,Berg2} and showed an impressive agreement with the
predictions of Ref.~\cite{bernard}. These experiments were extended to finite
photon four-momentum transfer $-Q^2$ via electro production at 
NIKHEF\cite{nikhef,Welch} and MAMI\cite{distler} at
$Q^2=0.1\,\mathrm{GeV^2/c^2}$, which is believed to be the limit of the
predictive power of HBChPT. Nevertheless, the results were in reasonable
agreement with the calculations\cite{BKMe96}.

In this paper we present a measurement at a value of
$Q^2=0.05\,\mathrm{GeV^2/c^2}$, half way between the photon point and the
existing electro production data. The present data cover the complete
azimuthal angle $\phi$ and allow for a Rosenbluth separation to extract the
unpolarized structure functions $\sigma_T$, $\sigma_L$, and
$\sigma_{TL}$. $\sigma_{TT}$ was observed to be so small that only upper
limits were obtained. Since the former experiments agree with the predictions
of HBChPT, the clear disagreement between our results and these calculations
are surprising.

\paragraph*{Formalism and Kinematics.}

In the one-photon exchange approximation, the unpolarized electro production
cross section can be written as (see e.g. Ref.~\cite{drechsel})
\newlength\nmln\settoheight\nmln{$\sqrt{2_L(1)}$}
\begin{eqnarray}
  \frac{d\sigma(\theta_\pi^*,\phi_\pi^*)}{d\Omega'dE'd\Omega_\pi^*} &=& 
  \Gamma \left(\rule{0mm}{\nmln}
    \sigma_T + \epsilon_L\,\sigma_{L} + \epsilon\,\sigma_{TT}\cos
    2\phi_\pi^*\right.\nonumber\\ 
  &&~~~~+\left.\sqrt{2\epsilon_L(1+\epsilon)}\,\sigma_{TL}\cos\phi_\pi^*\right)
  \label{equ:cross}
\end{eqnarray}
with the virtual photon flux $\Gamma$, photon energy $\omega^*$, and transverse
polarization $\epsilon$, $\epsilon_L=\epsilon Q^2/\omega^{*2}$. The pion
emission angles $\theta_\pi^*$ and $\phi_\pi^*$ are relative to the momentum
transfer {\boldmath$q$} and the scattering plane. Variables in the
photon-proton center of mass frame are denoted with an asterisk.

In the threshold region, where only $s$ and $p$ waves have to be taken into
account, the cross section can be further decomposed as
\begin{eqnarray}
  \sigma_{T}(\theta_\pi^*) & = 
  &({p_{\pi}^*}/{k_{\gamma}^*})\left(A + B \cos\theta_\pi^* +C
    \cos^2\theta_\pi^* \right), \nonumber \\
  \sigma_{L}(\theta_\pi^*)  & = 
  &({p_{\pi}^*}/{k_{\gamma}^*})\left(A' + B'\cos\theta_\pi^* +C'
    \cos^2\theta_\pi^* \right), \nonumber \\
  \sigma_{TL}(\theta_\pi^*)  & = 
&({p_{\pi}^*}/{k_{\gamma}^*})\left(D \sin\theta_\pi^* + E
    \sin\theta_\pi^*\cos\theta_\pi^* \right), \nonumber \\
  \sigma_{TT}(\theta_\pi^*)  & = 
&({p_{\pi}^*}/{k_{\gamma}^*})\left(F \sin^2\theta_\pi^*\right),
  \label{equ:multi}
\end{eqnarray}
where ${p_{\pi}^*}/{k_{\gamma}^*}$ is the ratio of pion CM momentum and photon
CM equivalent momentum. The angular coefficients are given by two $s$ wave
and five $p$ wave multipole combinations\cite{BKMe96}
\begin{eqnarray}
  A  & = & |E_{0+}|^2 + \frac{1}{2}\left(|P_2|^2+|P_3|^2\right),\nonumber\\
  B  & = & 2 \hbox{Re} \left(E_{0+}P_1^*\right),\nonumber\\
  C  & = & |P_1|^2 - \frac{1}{2}\left(|P_2|^2+|P_3|^2\right),\nonumber\\[1mm]
  D  & = & - \hbox{Re} \left(E_{0+} P_5^* + L_{0+} P_2^*\right),\nonumber\\
  E  & = & - \hbox{Re} \left(P_{1}P_{5}^* + P_4P_2^*\right),\nonumber\\
  F  & = & \frac{1}{2} \left( |P_2|^2 - |P_3|^2 \right),\nonumber\\[1mm]
  A' & = & |L_{0+}|^2 + |P_5|^2 ,\nonumber\\
  B' & = & 2 \hbox{Re} \left(L_{0+}P_4^*\right),\nonumber\\
  C' & = & \left(|P_4|^2 - |P_5|^2\right).
  \label{equ:coeff}
\end{eqnarray}

At threshold, the $s$ wave multipoles $E_{0+}$ and $L_{0+}$ are real and their
energy dependence is governed by the unitary cusp caused by the two step
$\gamma^*p \rightarrow \pi^{+} n \rightarrow \pi^{0}p$ amplitude\cite{aron}. 
The $p$
wave amplitudes $P_i$ are real and proportional to the pion center of mass
momentum $p_{\pi}^*$. The coefficient $F$ is too small to be extracted in this
experiment, so only the $p$ wave combination $P_{23}^{2}=(|P_2|^2+|P_3|^2)/2$
can be determined.

\paragraph*{Experiment.}

The experiment was performed at the three-spectrometer setup of the A1
Collaboration at MAMI (see Ref.~\cite{dreispek} for a detailed description). A
liquid Hydrogen target with a length of 49\,mm and target walls of 10\,$\mu$m
Havar was used with beam currents of up to $30\,\mu\mathrm{A}$, i.e.\ at a
luminosity of $\mathrm{39\,MHz/\mu b}$. 
For the detection of the
scattered electron, spectrometer B with 5.6\,msr solid angle at 15\% momentum
acceptance was used for the two forward angles, while spectrometer C with
22.5\,msr solid angle at 25\% momentum acceptance was used for the backward
angle.

For the detection of the recoil proton, spectrometer A with 21\,msr solid
angle and a momentum acceptance of 20\% was used for all settings.
Table~\ref{tab:settings} shows the central angle and momentum settings of the
spectrometers. Due to the kinematical focusing by the Lorentz boost, the full
center of mass solid angle was covered within each setting up to a center of
mass energy of $\Delta W = 4\,\mathrm{MeV}$ above threshold.

\begin{table}
  \caption{Kinematical settings for the central trajectories of the
    spectrometers. The four-momentum transfer for all settings is
    $Q^2=0.05\,\mathrm{GeV^2/c^2}$}
  \label{tab:settings}
  \begin{ruledtabular}
    \begin{tabular}{ccccccc}
    Spectr. & $\epsilon$ & $E_0$ & $E'$ &
    $p_p^{\mathrm{lab}}$ &$\theta_e^{\mathrm{lab}}$
    & $\theta_p^{\mathrm{lab}}$\\ && (MeV) & (MeV) & (MeV/c) &&\\
    \hline
    A$\wedge$B & 0.93 & 855 & 662.3 & 240.0 & $16.8^{\circ}$ &$44.6^{\circ}$\\
    A$\wedge$B & 0.72 & 435 & 263.7 & 233.1 & $38.6^{\circ}$ &$35.7^{\circ}$\\
    A$\wedge$C & 0.49 & 330 & 144.0 & 240.0 & $58.5^{\circ}$ &$28.7^{\circ}$\\
    \end{tabular}
  \end{ruledtabular}
\end{table}

All spectrometers were equipped with a detector package consisting of four
layers of vertical drift chambers for position and angle reconstruction, and
two layers of scintillator detectors for time-of-flight measurement and
trigger. For the electron arm, the trigger was defined by the coincidence of
both scintillator layers, while the low energy protons were already stopped in
the first scintillator layer of spectrometer A. A halocarbon gas \v{C}erenkov
detector with a detection efficiency of 99\% in this energy range was used for
$\pi^-$/$e^{-}$ separation, but not as part of the online trigger decision. The
electron detection efficiency and calibration was checked by a measurement of
elastic electron scattering.

\paragraph*{Analysis and Error Estimate.}

The coincidence time between electron and proton spectrometer, corrected for
the flight path in the spectrometers, was determined with a resolution of
$1.8\,\mathrm{ns}$ FWHM. This was limited by the uncertainty in the flight
path due to multiple scattering of the low energy protons in the detector
package of spectrometer A.

After a cut on the coincidence time, the missing mass was determined by the
four-momentum subtraction of incoming and outgoing particles, i.e. by
$m_{miss}^2 = (e_{in} + p_{in} - e_{out} - p_{out})^2$. A missing mass resolution of
$2.0\,\mathrm{MeV/c^2}$ was achieved, again limited by the multiple scattering
of the low energy protons.

The phase space was calculated in a Monte-Carlo simulation including
resolution, multiple scattering, energy loss, and radiative corrections, the
later based on the formulas of Ref.~\cite{MoTsai}.

The background contribution was determined by a cut in the coincidence time
spectrum. Since the modest missing mass resolution allowed only a loose cut of
$\pm 3\,\mathrm{MeV/c^2}$ around the $\pi^{0}$ mass, the signal to background
ratio was $\approx 1$ averaged over the acceptance. The background, however,
could be determined with a timing cut of 10 times the width of the cut for the
true events, leading to no significant contribution to the statistical error
by the background subtraction.

The systematic error is dominated by different effects at low and high
energies. For the lowest energy bin, the calibration of the momentum of the
electron arm is crucial, since assuming a too high electron momentum leads to a
smaller photon momentum and a smaller total CM energy, so that some of the
events are shifted below threshold. This error was estimated by varying the
central momentum of the electron arm in the analysis. For the lowest energy
bins we assumed $\delta\sigma/\sigma = 20\%$ error at $\Delta
W=0.5\,\mathrm{MeV}$ and 10\% at $\Delta W=1.5\,\mathrm{MeV}$.

For the higher energy bins, the acceptance of the proton arm is not uniform,
and a few angular bins are multiplied by large phase space correction
factors. This contribution to the systematic error can be estimated by varying
software cuts on the acceptance.

Compared to these two effects, the contribution of the efficiency and other
corrections can be neglected. Over all, we estimated an error of
$\delta\sigma/\sigma = 5\%~(3\%)$ at $\Delta
W=2.5\,\mathrm{MeV}~(3.5\,\mathrm{MeV})$.

After determining the two-fold cross section
$\sigma(\theta^*_{\pi},\phi^*_{\pi})$ for the three epsilon points, a fit with
the known $\phi^*_{\pi}$ dependence (Eqn.~\ref{equ:cross}) was performed in
each $\theta^*_{\pi}$ bin to extract $\sigma_0(\theta^*_{\pi}) =
\sigma_T(\theta^*_{\pi})+\epsilon_L\sigma_L(\theta^*_{\pi})$ and
$\sigma_{TL}(\theta^*_{\pi})$.

\begin{figure}
  \centering\includegraphics{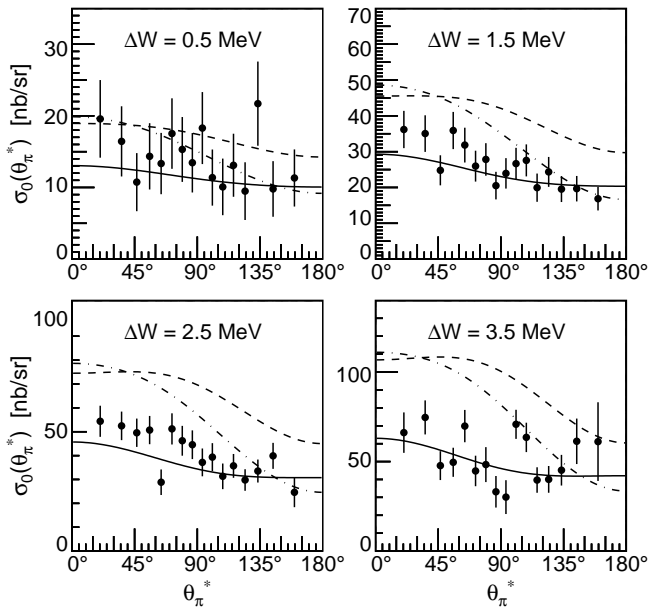}
  \caption{Differential cross sections for the first 4\,MeV above threshold
    for the virtual photon polarization $\epsilon=0.72$. The solid line
    represents a fit with the assumption of only $s$ and $p$ waves
    contributing, the dashed and dash-dotted lines represent the predictions of
    HBChPT{\protect\cite{BKMe96}} and MAID{\protect\cite{maid}}.}
\label{fig:cross072}
\end{figure}
\begin{figure}
  \centering\includegraphics{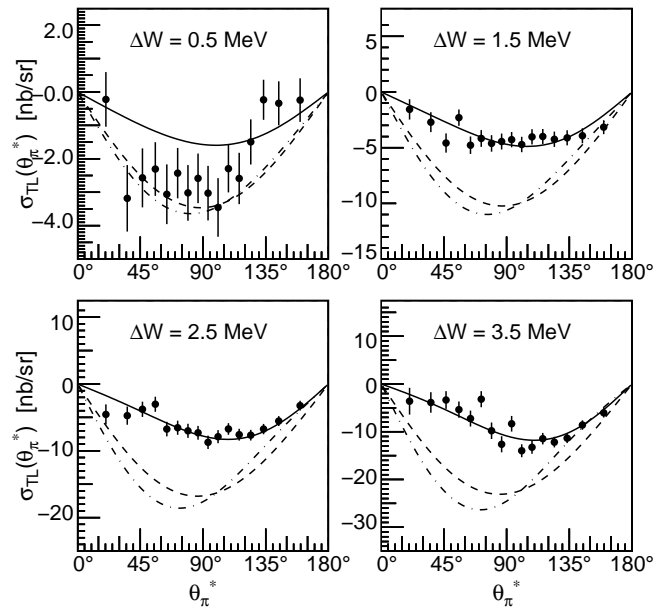}
  \caption{The transverse-longitudinal interference structure
    function, determined as weighted average of all three settings for
    $\epsilon$. Assignment of lines as in Fig.~\ref{fig:cross072}.}
  \label{fig:crossTL}
\end{figure}
\paragraph*{Results.}

Fig.~\ref{fig:cross072} shows the differential cross section for four energy
bins at the medium epsilon point in comparison with the predictions of
HBChPT\cite{BKMe96} and a phenomenological Model (MAID)\cite{maid}. The
complete data set can be obtained from our web site. Fig.~\ref{fig:crossTL}
shows the transverse-longitudinal interference structure function extracted as
weighted average from all three settings.

In order to extract the multipole amplitudes, a fit using Eqn.~\ref{equ:coeff}
with the assumption of constant real $s$ wave multipoles
(i.e. $E_{0+}$, $L_{0+}$ constant in energy) and real $p$ wave multipoles
proportional to the pion CM momentum ($P_i=p_{\pi}^*\hat{P}_i$ with
$\hat{P}_i$ constant in energy) was performed, indicated by a solid line in
the figures. These assumptions are not exactly valid, since a variation of
$\approx 5\%$ in the total $s$ wave amplitude is expected due to the unitary
cusp\cite{BKMe96,aron}, but the statistical significance of the data is not
sufficient to resolve this variation. The extracted fit parameters are
presented in the first row of table~\ref{tab:fitparameter}.

It is important to note, that a least squares fit cannot give a real
picture of the sensitivity of the data to the multipole amplitudes. The data
set is dominated by the systematic error and thus violates a fundamental
precondition of Gaussian distributed errors for a $\chi^2$ fit.

\begin{table}
  \caption{Extracted multipole amplitudes in comparison with the
    threshold values of HBChPT{\protect\cite{BKMe96}} and
    MAID{\protect\cite{maid}}. The AmPS{\protect\cite{Welch}} value for
    $|L_{0+}|$ was extracted from their value for $a_{0}\approx\epsilon_{L}
    |L_{0+}|^{2}$. For the AmPS {\protect\cite{nikhef}} fit $L_{0+}$ was
    fixed, since no Rosenbluth separation was performed.}
  \label{tab:fitparameter}
  \begin{ruledtabular}
  \begin{tabular}{ldddddd}
    &\multicolumn{1}{c}{$E_{0+}$} 
    &\multicolumn{1}{c}{$L_{0+}$}
    &\multicolumn{1}{c}{$\hat{P_{23}}^2$} 
    &\multicolumn{1}{c}{$\hat{P_1}$}
    &\multicolumn{1}{c}{$\hat{P_4}$} 
    &\multicolumn{1}{c}{$\hat{P_5}$} \\
    &\multicolumn{2}{c}{$\left(10^{-3}m_{\pi}^{-1}\right)$}
    &\multicolumn{1}{c}{\hspace{-2mm}
                        $\left(10^{-6}m_{\pi}^{-4}\right)$}
    &\multicolumn{3}{c}{$\left(10^{-3}m_{\pi}^{-2}\right)$}\\
    \hline ~\\[-1mm]
    \multicolumn{7}{c}{$Q^2=0.05\,\mathrm{GeV^2/c^2}$}\\[1mm]
    Fit   &    0.57 &   -1.29 &   100 &     12.0  &    0.29 &    -1.9 \\
    Error & \pm0.11 & \pm0.02 &  \pm3 &   \pm0.3  & \pm0.33 &  \pm0.3 \\
    AmPS{\protect\cite{Welch}}
          &         & (-)1.57 &       &           &         &         \\
          &         & \pm0.96 &       &           &         &         \\
    ChPT  &    0.27 &   -1.55 &   353 &     16.5  &   -0.72 &    -0.2 \\
    MAID  &    0.76 &   -1.4  &   250 &     15.0  &   -1.75 &     1.9 \\[4mm]
    \multicolumn{7}{c}{photon point  $Q^2=0\,\mathrm{GeV^2/c^2}$}     \\[1mm]
    MAMI{\protect\cite{Schmidt}}  
          &   -1.33 &         &   111 &      9.5  &         &         \\
    ChPT  &   -1.14 &   -1.70 &   105 &      9.3  &   -0.6  &    -0.2 \\
    MAID  &   -1.16 &   -1.29 &    95 &      9.3  &   -3.0  &     2.2 \\[4mm]
    \multicolumn{7}{c}{$Q^2=0.1\,\mathrm{GeV^2/c^2}$}                 \\[1mm]
    MAMI{\protect\cite{distler}}
          &    0.58 &   -1.38 &   573 &    15.1   & -2.3    &     0.1 \\
          & \pm0.18 & \pm0.01 & \pm11 &  \pm0.8   &  \pm0.2 &  \pm0.3 \\
    AmPS{\protect\cite{nikhef}}
          &    1.99 &   -1.33 &   526 &    16.4   &   -1.0  &    -1.0 \\
          & \pm0.3  &\multicolumn{1}{c}{fixed}
                              &\pm7&  \pm0.6   &  \pm0.4 &  \pm0.4 \\
    ChPT  &    1.42 &   -1.33 &   571 &     20.1  &   -0.6  &    -0.1 \\
    MAID  &    2.2  &   -1.12 &   315 &     17.1  &   -1.1  &     1.4 \\
  \end{tabular}
  \end{ruledtabular}
\end{table}

\begin{figure}
  \centering\includegraphics{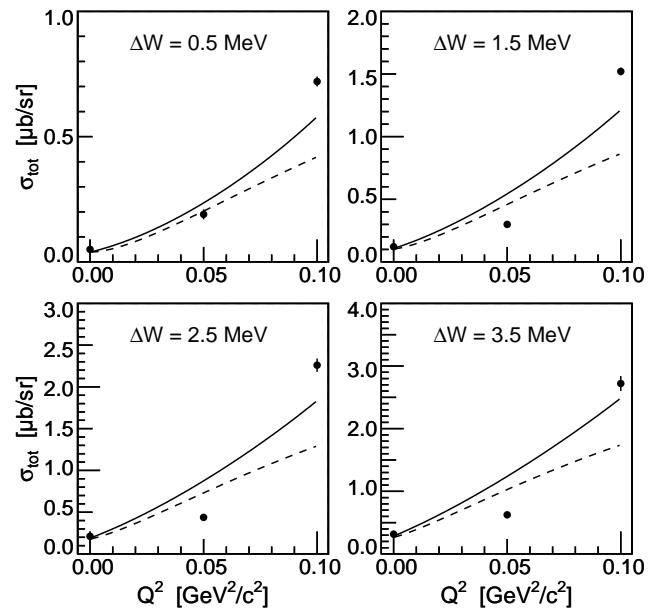}
  \caption{The total cross section $\sigma_{tot}$ versus $Q^2$, 
    at a value of $\epsilon = 0.8$. The solid (dashed)
    line is the prediction of ChPT (MAID), data points at $Q^2=0$ and
    $0.1\,\mathrm{GeV^2/c^2}$ from {\protect\cite{Schmidt,distler}}.}
  \label{fig:crossqsq}
\end{figure}

To compare the present result with former experiments in a consistent way,
the same fit was performed to the data sets at $Q^2 = 0.1\,\mathrm{GeV^2/c^2}$,
which are believed to be more or less consistent with HBChPT on the cross
section level. The fit results are included in table~\ref{tab:fitparameter}
and differ from the multipole amplitudes quoted in the corresponding
references\cite{distler,nikhef}, which were extracted using model assumptions.

Fig.~\ref{fig:crossqsq} shows the unpolarized total cross section
$\sigma_{tot}=\int (\sigma_T+\epsilon_L\sigma_L) d\Omega$ as a function of
$Q^2$ to further illustrate the discrepancy between the currently available
data and the calculations in a model independent manner. While the photo
production data could be described by both, HBChPT and MAID, the strongest
deviation appears at the $Q^2$ value of the present experiment, while already
at $Q^2=0.1\,\mathrm{GeV^2/c^2}$, where the parameters of HBChPT were
determined, a deviation from the HBChPT calculation of up to 20\% appears. The
disagreement with the MAID model is even larger.

From the fit parameters quoted in table~\ref{tab:fitparameter} one can see,
that the major part of the discrepancy comes from the large $p$ wave multipole
combination $P_{23}^2$. This is a serious problem for the prediction of
HBChPT, since this combination is already fixed by the photoproduction data
and cannot be adjusted by varying the free parameters of the calculation to
the current order. The strong curvature in the $Q^2$ dependence of
Fig.~\ref{fig:crossqsq} clearly indicates, that higher orders in $Q^2$ have to
be included in the $p$ wave description.

In order to check the surprising result of our experiment, we repeated our
differential cross section measurement for the highest epsilon (0.92) in an
independent experiment. The result of this experiment agreed with the present
experiment within the errors. In addition, since these results rely on the
quality of the simulation and phase space integration program, a new,
completely independent code was written to check the phase space integration.

In summary, it appears that there is a significant discrepancy between HBChPT
and phenomenological models on the one hand and the experimental data on the
other hand. However, an inconsistency in the different electro production data
sets, which were all taken in separate experiments, can not be excluded. As
mentioned above, a repeated measurement at $Q^2=0.05\,\mathrm{GeV^2/c^2}$ and
$\epsilon=0.92$ at MAMI confirmed our result. We plan to further explore the
$Q^2$ dependence in a future experiment. An independent measurement at the
Jefferson Lab in the same $Q^2$ range is also planned\cite{jlab}.

\begin{acknowledgments}
This work was supported by the Deutsche Forschungsgemeinschaft (SFB~443), and
the Federal State of Rhine\-land-Palatinate.  A.\,M.~Bernstein is grateful to
the Alexander von Humboldt Foundation for a Research Award.
\end{acknowledgments}

\end{document}